\documentclass[a4paper,fleqn]{cas-dc}

\usepackage[numbers]{natbib} 

\usepackage{graphics}
\usepackage{bbold}
\usepackage{color}

\usepackage{multirow}

\usepackage{array}
\usepackage{booktabs}
\usepackage{amsmath}
\usepackage{graphicx}
\usepackage{subfigure}

\usepackage{lineno}

\usepackage{amsmath}
\usepackage{slashed}

\usepackage{lineno,hyperref}

\def\be		{\begin{eqnarray}}
\def\en		{\end{eqnarray}}
\def\nen	{\nonumber\end{eqnarray}}
\def\no		{\nonumber}

\def\ug		{\ensuremath{\!\!\!\!\!&=&\!\!\!\!\!}}
\def\jp		{\ensuremath{J\!/\psi}}
\def\psii		{\ensuremath{\psi(2S)}}

\def\ee		{\ensuremath{e^+e^-}}

\def\bb		{\ensuremath{\mathcal{B}\overline{\mathcal{B}}}}
\def\b		{\ensuremath{\mathcal{B}}}

\def \br {{\rm{BR}}}

\begin{document}

\shorttitle{Amplitudes separation and strong-electromagnetic relative phase in the $\psi(2S)$ decays into baryons}  
\title[mode = title]{Amplitudes separation and strong-electromagnetic relative phase in the $\psi(2S)$ decays into baryons}               

\shortauthors{R.~Baldini Ferroli et al.}

\author[1]{Rinaldo~Baldini~Ferroli}
\address[1]{INFN Laboratori Nazionali di Frascati, I-00044, Frascati, Italy}
\author[2]{Alessio~Mangoni}
\author[2,3]{Simone~Pacetti}
\address[2]{INFN Sezione di Perugia, I-06100, Perugia, Italy}
\address[3]{Universit\`a di Perugia, I-06100, Perugia, Italy}
\author[4]{Kai~Zhu}
\address[4]{Institute of High Energy Physics, I-100049, Beijing People's Republic of China}

\begin{abstract}
The strong, electromagnetic and mixed strong-electromagnetic amplitudes of the $\psi(2S)$ decays into baryon-anti-baryon pairs have been obtained by exploiting all available data sets in the framework of an effective Lagrangian model. 
\\
 We observed that at the $\psi(2S)$ mass the QCD regime is not completely perturbative, as can be inferred by the relative strength of the strong and the mixed strong-electromagnetic amplitudes. Recently a similar conclusion has been reached also for the $J/\psi$ decays.
 \\
 The relative phase between the strong and the electromagnetic amplitudes is $\varphi = (58\pm 8)^\circ$, to be compared with $\varphi = (73\pm 8)^\circ$ obtained for the $J/\psi$. 
 \\
 On the other hand, in the case of the $\psi(2S)$ meson, different values of the ratio between strong and mixed strong-electromagnetic amplitudes are phenomenologically required, while for the $J/\psi$ meson only one ratio was enough to describe the data.
 \\
   Finally, we also observed a peculiar behavior of the mixed strong-electromagnetic amplitudes of the decays $\psi(2S)\to\Sigma^+ \overline \Sigma{}^-$ and $\psi(2S)\to\Sigma^- \overline \Sigma{}^+$. 
\end{abstract}


\maketitle

\section{Introduction}%
Since their discovery, the $c \overline c$ mesons, the so-called {\it charmonia}, have been representing unique tools to expand our knowledge on the dynamics of the strong interaction at various energy ranges. The hadronic decays of the $J/\psi$ meson, a charmonium with quantum numbers $I^G(J^{PC}) = 0^-(1^{--})$, mass $M_{\jp} \simeq 3.1 \ \rm GeV$ and width $\Gamma_{\jp} \simeq 9.3~\cdot~10^{-5}$~GeV~\cite{Tanabashi:2018oca}, have been deeply investigated. Recently, it has been found that they occur halfway between the perturbative and non-perturbative QCD regime~\cite{Ferroli:2019nex}. Moreover, it has been shown that the mixed strong-electromagnetic (strong-EM) amplitude of the $J/\psi$ decays is not always negligible~\cite{Ferroli:2016jri}. 
\\
The procedure to single out the strong, the EM and the mixed strong-EM amplitudes of the decay of a charmonium state into baryon-anti-baryon ($\mathcal B \overline{\mathcal B}$) pairs belonging to the spin-1/2 SU(3) baryon octet has been defined and implemented for the first time in the case of the \jp\ meson~\cite{Ferroli:2019nex}.
\\
Such a procedure is based on an effective Lagrangian containing SU(3) symmetry breaking terms, depending on a set of coupling constants to be determined by means of a $\chi^2$ minimization.
\\
The whole data sets available in the PDG~\cite{Tanabashi:2018oca}, together with new results provided by the BESII Collaboration~\cite{Ablikim:2017tys} have been used.
\\
In the case of the \jp, the strong, the EM and the mixed strong-EM contributions to the total  branching ratio (BR), as well as a strong-EM relative phase of $(73\pm 8)^\circ$ have been determined~\cite{Ferroli:2019nex}.
\\
 The $\psii$ meson is a charmonium with the same quantum numbers of the $\jp$ meson, i.e., $I^G(J^{PC}) = 0^-(1^{--})$, mass $M_{\psii} \simeq 3.7 \ \rm GeV$ and width $\Gamma_{\psii} \simeq 2.9 \cdot 10^{-4} \ \rm GeV $~\cite{Tanabashi:2018oca}. The $\psii \to \bb$ decays, where $\mathcal B$ is a spin-1/2 baryon of the SU(3) octet represented by the matrix
\be
B=\begin{pmatrix}
\Lambda/\sqrt{6}+\Sigma^0/\sqrt{2} & \Sigma^+ & p\\
\Sigma^- & \Lambda/\sqrt{6}-\Sigma^0/\sqrt{2} & n\\
\Xi^- & \Xi^0 & -2 \Lambda/\sqrt{6}\end{pmatrix} \,,
\nen
can be studied by means of the same procedure developed and successfully implemented in the case of the $\jp$ meson.
\\
However, even though in the framework of the same procedure and despite their common nature of charmonia, the decay mechanisms of the two mesons \jp\ and \psii\ have different characteristics. This is also proven by the different angular distributions of the final baryons in the processes $\ee\to\jp \to \Sigma^0 \overline \Sigma{}^0$ and $\ee\to\psii \to \Sigma^0 \overline \Sigma{}^0$, observed by the BESIII Collaboration~\cite{Ablikim:2017tys} and discussed in Ref.~\cite{Ferroli:2018yad}.
\section{Effective Lagrangian}
The effective Lagrangian $\mathcal L$ used for the $\psii$ .decays into baryon-antibaryon pairs can be written as~\cite{Ferroli:2019nex}
\be
\mathcal L \propto {\rm Tr}(B \overline B) + \mbox{[symmetry breaking terms]} \,,
\nen
where the SU(3) symmetry breaking terms are due to EM and quark mass difference effects. The EM breaking effects depend on the quark coupling with the photon given by
\be
\overline q \gamma^\mu \Lambda_{\rm E} q \equiv {2 \over 3} \overline u \gamma^\mu u - {1 \over 3} \overline d \gamma^\mu d - {1 \over 3} \overline s \gamma^\mu s \,,
\nen
so that, the matrix $\Lambda_E$ turns out to be the following combination of the third and the eighth Gell-Mann matrices
\be
\Lambda_{\rm E} = {1 \over 2} \left( \lambda_3 + {\lambda_8 \over \sqrt 3} \right) \,.
\nen
Similarly, the quark mass difference breaking effects are related to the mass term
\be
\overline q \Lambda_{\rm M} q \equiv m_u \overline u u +m_d \overline d d+m_s \overline s s \,,
\nen
and the corresponding matrix $\Lambda_M$, in terms of the Gell-Mann and the three-dimensional identity matrices, has the form
\be
\Lambda_{\rm M} = m_0 I_3 + {m_d-m_s \over \sqrt 3} \lambda_8 + {m_u-m_d \over 6} \big(2 I_3+3\lambda_3+\sqrt 3\lambda_8 \big) \,,
\nen
where
\be
m_0 = {m_u+m_d+m_s \over 3} \,.
\nen
We keep the SU(2) symmetry exact, assuming that $m_u = m_d$, so that
\be
\Lambda_{\rm M} = m_0 I_3 + {m_d-m_s \over \sqrt 3} \lambda_8 \,, \ \ \ \ \ m_0 = {2m_d-m_s \over 3} \,.
\nen
Therefore, the SU(3) symmetry breaking terms are related to the so-called spurion matrices
\be
S_e = g_e \Lambda_{\rm E} \,, \ \ \ \ \ S_m = g_m \big( \Lambda_{\rm M} - m_0 I_3 \big) \,,
\nen
where $g_e$ and $g_m$ are the coupling constants. The full Lagrangian is obtained by adding to the leading term proportional to ${\rm Tr}(B \overline B)$, additional terms proportional to~\cite{Zhu:2015bha}
\be
{\rm Tr}\big(\{B, \overline B\} S\big) \,, \ \ \ \ \ {\rm Tr}\big([B, \overline B] S\big) \,,
\nen
where $S \in \{ S_e,S_m \}$. 
\section{Amplitudes separation}
The amplitude for the decay $\psii \to \bb$, where $\bb$ is a baryon-antibaryon pair of the SU(3) octet, can be written as
\be
\mathcal A_{\bb}=\mathcal A^{ggg}_{\bb} + \mathcal A^{gg\gamma}_{\bb} + \mathcal A^\gamma_{\bb} \,,
\nen
\begin{figure}
\centering
	\includegraphics[width=.6\linewidth]{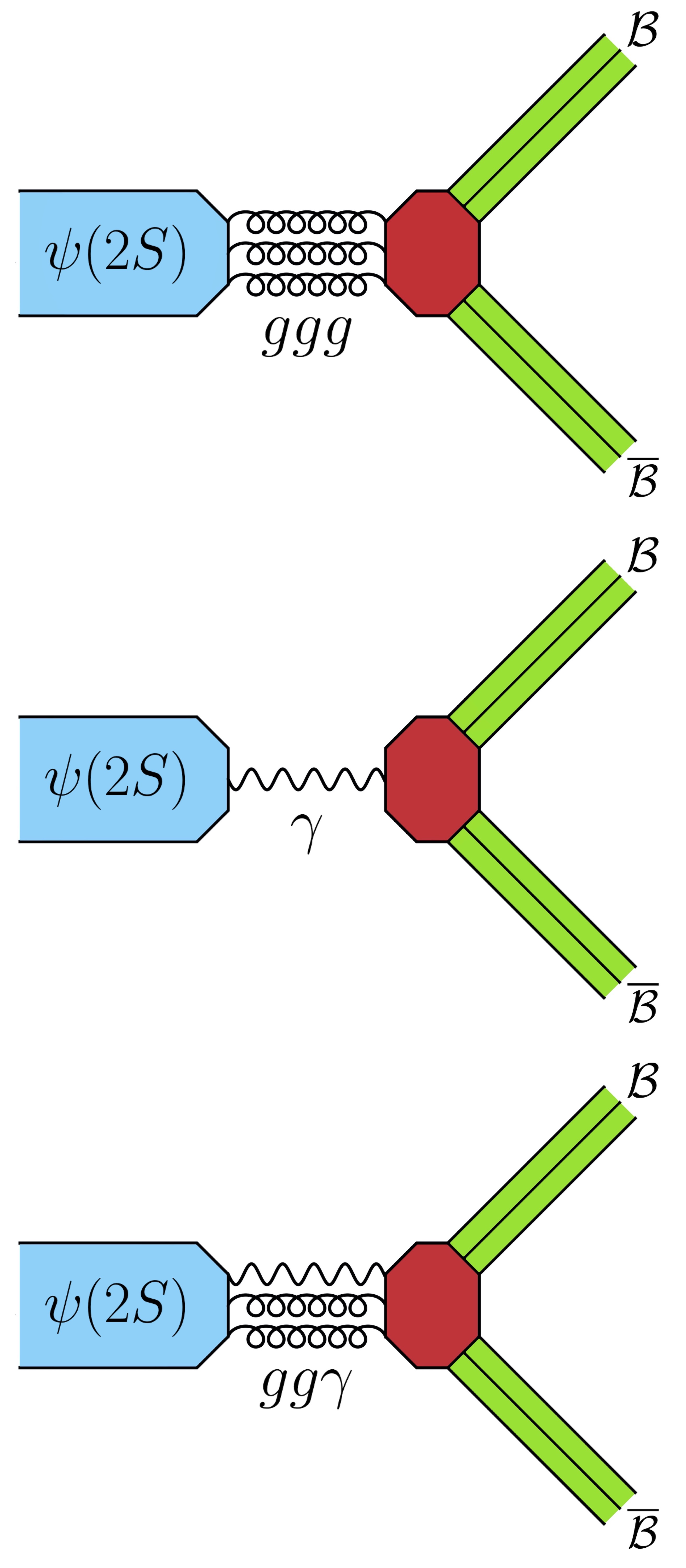}
	\caption{\label{fig:feyn}Feynman diagrams of the strong EM and mixed strong-EM contributions for the decay $\psi(2S) \to \bb$.}
\end{figure}%
where $\mathcal A^{ggg}_{\bb}$, $\mathcal A^\gamma_{\bb}$ and $\mathcal A^{gg \gamma}_{\bb}$ are the strong, the EM and the mixed strong-EM amplitudes, related to the Feynman diagrams shown in Fig.~\ref{fig:feyn}. 
\\
Following Ref.~\cite{Ferroli:2019nex}, we parametrize the three amplitudes for each final state as shown in Table~\ref{tab:A.par.pre}. The parameter $G_0$ is related to the SU(3) exact symmetry; $D_m$ and $F_m$ account for the quark mass difference breaking term; $D_e$ and $F_e$ for the EM one; $\varphi$ is the relative phase between strong and EM amplitudes.
\\
The mixed strong-EM amplitudes is vanishing for the decays into neutranl particles~~\cite{Chernyak:1984bm,Claudson:1981fj,Ferroli:2019nex}.
\\
 Due to the peculiarity of the \psii\ meson we consider two ratios between mixed strong-EM and strong amplitudes,  instead of a single one as in the \jp\ case. Indeed, we separate the four decays with a non-zero mixed contribution in two groups that differ by the number of strong correction parameters, namely
\be
\begin{aligned}
\bb=p \overline p,\, \Xi^- \overline \Xi{}^+ \ \ \to \ \ \mathcal A^{ggg}_{\bb} \propto G_0 - D_m \pm F_m \ \ \to \ \ R_1 \,, \\
\bb=\Sigma^+ \overline \Sigma{}^-,\, \Sigma^- \overline \Sigma{}^+ \ \ \to \ \ \mathcal A^{ggg}_{\bb} \propto G_0 + 2 D_m  \ \ \to \ \ R_2 \,,
\label{eq.choice.2R}
\end{aligned}
\en
hence $R_1$ and $R_2$ are the ratios
\be
R_{1} = {\mathcal A^{gg \gamma}_{p \overline p} \over \mathcal A^{ggg}_{p \overline p}} = {\mathcal A^{gg \gamma}_{\Xi^- \overline \Xi{}^+} \over \mathcal A^{ggg}_{\Xi^- \overline \Xi{}^+}} \,, \ \ \ \ \ R_{2} = {\mathcal A^{gg \gamma}_{\Sigma^+ \overline \Sigma{}^-} \over \mathcal A^{ggg}_{\Sigma^+ \overline \Sigma{}^-}} = {\mathcal A^{gg \gamma}_{\Sigma^- \overline \Sigma{}^+} \over \mathcal A^{ggg}_{\Sigma^- \overline \Sigma{}^+}} \,.
\nen
\begin{table}
\vspace{-2mm}
\centering
\caption{Parameterizations of the \bb\ decay amplitudes.}
\label{tab:A.par.pre} 
\begin{tabular}{llll} 
\hline\noalign{\smallskip}
\bb & $\mathcal A^{ggg}_{\bb}$ & $\mathcal A^{gg\gamma}_{\bb}$ & $\mathcal A^\gamma_{\bb}$ \\
\noalign{\smallskip}\hline\noalign{\smallskip}
$\Sigma^0 \overline \Sigma{}^0$ & $(G_0 + 2 D_m) e^{i \varphi}$ & $0$ & $D_e$ \\
$\Lambda \overline \Lambda$ & $(G_0 - 2 D_m) e^{i \varphi}$ & $0$ & $- D_e$ \\
$\Lambda \overline \Sigma{}^0+$ c.c. & $0$ & $0$ & $\sqrt{3}\,D_e$ \\
$p \overline p$ & $(G_0 - D_m + F_m) e^{i \varphi}$ & $\mathcal A^{ggg}_{p \overline p}R_1$ & $D_e + F_e$ \\
$n \overline n$ & $(G_0 - D_m + F_m) e^{i \varphi}$ & $0$ & $- 2\,D_e$ \\
$\Sigma^+ \overline \Sigma{}^-$ & $(G_0 + 2 D_m) e^{i \varphi}$ & $\mathcal A^{ggg}_{\Sigma^+ \overline \Sigma{}^-}R_2$ & $D_e + F_e$ \\
$\Sigma^- \overline \Sigma{}^+$ & $(G_0 + 2 D_m) e^{i \varphi}$ & $\mathcal A^{ggg}_{\Sigma^- \overline \Sigma{}^+}R_2$ & $D_e - F_e$ \\
$\Xi^- \overline \Xi{}^+$ & $(G_0 - D_m - F_m) e^{i \varphi}$ & $\mathcal A^{ggg}_{\Xi^- \overline \Xi{}^+}R_1$ & $D_e - F_e$ \\
$\Xi^0 \overline \Xi{}^0$ & $(G_0 - D_m - F_m) e^{i \varphi}$ & $0$ & $- 2\,D_e$ \\
\noalign{\smallskip}\hline
\end{tabular}
\end{table}%
This choice is also phenomenologically suggested by the data as we will see later.
\\
 Asymptotically these ratios are real and, once normalized by the electric charge, have the same value, as predicted by the perturbative QCD (pQCD)~\cite{Korner:1986vi}
\be
|R_{\rm pQCD} (q^2)| \mathop{\sim}_{q^2\gg\Lambda^2_{\rm QCD}} {4 \over 5} {\alpha \over \alpha_S(q^2)} \,.
\label{eq:RpQCD}
\en
Since at the $\psii$ mass, as well as at the $\jp$ mass~\cite{Ferroli:2019nex}, the regime cannot be considered completely perturbative, we treat $R_1$ and $R_2$ as free parameters to be determined by means of a fitting procedure.
\\
In our model we assume that, for the SU(3) octet, there is only one relative phase, $\varphi$, among the strong and the EM amplitudes. Moreover, the strong and the mixed amplitudes are assumed to be relatively real, i.e., also the two ratios $R_1$ and $R_2$ are supposed to be real.
\section{Electromagnetic couplings and $\chi^2$ definition}
The cross section for the annihilation $e^+e^- \to p \overline p$ at the $\psii$ mass can be calculated using the following formula obtained by the BESIII Collaboration~\cite{Ablikim:2019njl}
\be
\sigma_{\ee\to p \overline p} \left(M_{\psii}^2 \right)
&\!\!\!=\!\!\!& {6912 \,\pi \alpha^2 \left(M_{\psii}^2\!\!+\!2M_p^2 \right) \over M_{\psii}^{12}\,{\rm GeV^{-8}}} \no \\
&\!\!\! \!\!\! & \cdot\left[ \ln^2\left(\!{M_{\psii}^2 \over 0.52^2\,\rm GeV^2}\!\right)\!+\!\pi^2 \right]^{\!\!-2} .
\nen
Using the obtained value it is possible to calculate the EM BR for the decay $\psii \to p \overline p$ by~\cite{Ferroli:2016jri}
\be
\label{eq.BRgamma}
\br^\gamma_{p \overline p} = \br_{\mu\mu} \, {\sigma_{\ee \to p \overline p}\left(M_{\psii}^2\right) \over \sigma^0_{\ee \to \mu^+ \mu^-}\left(M_{\psii}^2\right)} \,,
\en
where $\br_{\mu\mu}$ is the BR of the decay $J/\psi \to \mu^+\mu^-$, and $\sigma^0_{\ee \to \mu^+ \mu^-}(q^2)$ represents the bare $\ee \to \mu^+ \mu^-$ cross section 
\be
\sigma^0_{\ee \to \mu^+ \mu^-} (q^2)= {4 \pi \alpha^2 \over 3 q^2} \,.
\nen
By using the PDG value~\cite{Tanabashi:2018oca} $\br_{\mu\mu} = (8.0 \pm 0.8) \times 10^{-6}$, we obtain
\be
\label{eq.BRgammapp}
\br^\gamma_{p \overline p} = (2.2 \pm 1.5) \times 10^{-6} \,,
\en
reported in the last row of Table~\ref{tab:BRJpsi.data}.\\
We define the $\chi^2$ function referred to the set of baryons for which there are available data, see Table~\ref{tab:BRJpsi.data}, as
\be
\label{eq.chi.2}
\chi^2\left(\xi\right) = \sum_{\bb} \left({\br_{\bb}^{\rm th}-\br_{\bb}^{\rm exp} \over \delta \br_{\bb}^{\rm exp}}\right)^2 + \left({\br_{p\overline{p}}^{\gamma,\rm th}-\br_{p\overline{p}}^{\gamma} \over \delta \br_{p\overline{p}}^{\gamma}}\right)^2 \,,
\en
where $\xi$ is the set of parameters
\be
\xi=\{G_0,D_e,D_m,F_e,F_m,R_1,R_2,\varphi \}\,,
\nen
and the sum runs over the baryon pairs of the set
\be
\bb\!=\! \left\{ \!\Sigma^0 \overline \Sigma{}^0, \Lambda \overline \Lambda, p \overline p, n \overline n, \Xi^0 \overline \Xi{}^0, \Xi^- \overline \Xi{}^+, \Lambda \overline \Sigma{}^0\!+\!{\rm c.c.}, \Sigma^+ \overline \Sigma{}^- \!\right\} \,.
\nen
The minimization is performed with respect to the eight parameters of the set $\xi$ using the data reported in Table~\ref{tab:BRJpsi.data}. The theoretical BRs, $\br_{\bb}^{\rm th}$ and $\br_{p\overline{p}}^{\gamma,\rm th}$, are given by combinations of these parameters as shown in Table~\ref{tab:A.par.pre}.
\begin{table}
\vspace{-2mm}
\centering
\caption{Branching ratios data from PDG~\cite{Tanabashi:2018oca}. The last two rows are from BESIII experiment~\cite{Ablikim:2016sjb,Ablikim:2019njl}.}
\label{tab:BRJpsi.data} 
\begin{tabular}{lrr} 
\hline\noalign{\smallskip}
Decay process & Branching ratio & Error \\
\noalign{\smallskip}\hline\noalign{\smallskip}
$\psi(2S) \to \Sigma^0 \overline \Sigma{}^0$ & $(2.35 \pm 0.09) \times 10^{-4}$ & $3.83 \%$ \\
$\psi(2S) \to \Lambda \overline \Lambda$ & $(3.81 \pm 0.13) \times 10^{-4}$ & $3.41 \%$ \\
$\psi(2S) \to \Lambda \overline \Sigma{}^0 + {\rm c.c.} $ & $(1.23 \pm 0.24) \times 10^{-5}$ & $19.5 \%$ \\
$\psi(2S) \to p \overline p$ & $(2.94 \pm 0.08) \times 10^{-4}$ & $2.72 \%$  \\
$\psi(2S) \to n \overline n$ & $(3.06 \pm 0.15) \times 10^{-4}$ & $4.90 \%$ \\
$\psi(2S) \to \Sigma^+ \overline \Sigma{}^-$ & $(2.32 \pm 0.12) \times 10^{-4}$ & $5.17 \%$ \\
$\psi(2S) \to \Xi^- \overline \Xi{}^+$ & $(2.87 \pm 0.11) \times 10^{-4}$ & $3.83 \%$  \\
$\psi(2S) \to \Xi^0 \overline \Xi{}^0$ & $(2.73 \pm 0.13) \times 10^{-4}$ & $4.76 \%$ \\
$\psi(2S) \to \gamma \to p \overline p$ & $(2.2 \pm 1.5) \times 10^{-6}$ & $68.2 \%$ \\
\noalign{\smallskip}\hline
\end{tabular}
\end{table}\\
\section{$\psi(2S)$ results}
\label{sec:results}
%
The results of the $\chi^2$ minimization are reported in Table~\ref{tab:results.par}, the errors have been obtained by means of a Monte Carlo procedure. For comparison we have reported also, in Table~\ref{tab:results.par-j}, the related results obtained for the $\jp$ meson, where the 1-$R$ hypothesis, i.e., $R=R_1=R_2$, was in good agreement with the data. 
\begin{table}
\vspace{-2mm}
\centering
\caption{Best values of the parameters, for the $\psii$ meson, describing the decay \bb\ amplitudes, see Table~\ref{tab:A.par.pre}, obtained by minimizing the $\chi^2$ defined in Eq.~\eqref{eq.chi.2}, using the data reported in Table~\ref{tab:BRJpsi.data}.}
\label{tab:results.par} 
\begin{tabular}{lr} 
\hline\noalign{\smallskip}
$\chi^2/N_{\rm dof}$ & $0.035$ \\
$G_0$ & $(4.580 \pm 0.060) \times 10^{-3} \ \rm GeV$ \\
$D_e$ & $(5.37 \pm 0.52) \times 10^{-4} \ \rm GeV$ \\
$D_m$ & $(-3.95 \pm 0.49) \times 10^{-4} \ \rm GeV$ \\
$F_e$ & $(-1.84 \pm 0.70) \times 10^{-4} \ \rm GeV$  \\
$F_m$ & $(-1.13 \pm 0.50) \times 10^{-4} \ \rm GeV$ \\
$\varphi$ & $1.02 \pm 0.15 = (58 \pm 8)^\circ $  \\
$R_1$ & $(-15.3 \pm 2.8) \times 10^{-2}$ \\
$R_2$ & $(2.1 \pm 4.1) \times 10^{-2}$ \\
\noalign{\smallskip}\hline
\end{tabular}
\end{table}%
\begin{table}
\vspace{-2mm}
\centering
\caption{Best parameters obtained for the $\jp$ meson under the 1-$R$ hypothesis~\cite{Ferroli:2019nex}.}
\label{tab:results.par-j} 
\begin{tabular}{lr} 
\hline\noalign{\smallskip}
$\chi^2/N_{\rm dof}$ & $1.33$ \\
$G_0$ & $(5.73511 \pm 0.0059) \times 10^{-3} \ \rm GeV$ \\
$D_e$ & $(4.52 \pm 0.19) \times 10^{-4} \ \rm GeV$ \\
$D_m$ & $(-3.74 \pm 0.34) \times 10^{-4} \ \rm GeV$ \\
$F_e$ & $(7.91 \pm 0.62) \times 10^{-4} \ \rm GeV$  \\
$F_m$ & $(2.42 \pm 0.12) \times 10^{-4} \ \rm GeV$ \\
$\varphi$ & $1.27 \pm 0.14 = (73 \pm 8)^\circ $  \\
$R$ & $(-9.7 \pm 2.1) \times 10^{-2}$ \\
\noalign{\smallskip}\hline
\end{tabular}
\end{table}\\%
The BRs are reported in Table~\ref{tab:BRcalc.psi2s}, together with the corresponding experimental values.  The obtained value for the BR of the unobserved decay $\psii~\to~\Sigma^-\overline\Sigma{}^+$ represents a prediction of the model.
\\
The minimum normalized $\chi^2$ is
\be
\frac{\chi^2\big(\xi^{\rm best}\big)}{N_{\rm dof}} = 0.035\,,
\label{eq:chi2}\en
where the number of degrees of freedom is $N_{\rm dof}=N_{\rm const}-N_{\rm param}=1$, having nine constraints, $N_{\rm const}=9$, and eight free parameters, $N_{\rm param}=8$. 
\\
The significance of the different mixed-to-strong amplitude ratios in the description of the \psii\ decay mechanism can be verified by comparing the normalized $\chi^2$'s, obtained in the case where $R_1$ and $R_2$ are considered as free parameters, Eq.~\eqref{eq:chi2}, to that in which they are considered equal each other, $R_1=R_2=R$. The minimum normalized $\chi^2$ is
\be
\frac{\chi^2\big(\mathcal{\xi'}^{\rm best}\big)}{N'_{\rm dof}}=\frac{14.32}{2} = 7.16\,,
\label{eq:chi2-2}\en
with the single ratio
\be
\label{eq-R.1R.S}
R=-0.077 \pm 0.029 \,
\en
and where
\be
\xi'=\{G_0',D_e',D_m',F_e',F_m',R,\varphi' \}\,,
\nen
is the set of the best values of the parameters obtained in this case, with $N'_{\rm dof} = N_{\rm dof}+1=2$. 
\\
Despite the quite low number of degrees of freedom we obtain the $p$-values
\be
p(0.035;1)= 0.852 \,, \ \ \ \ \ p(14.32;2)= 7.77 \times 10^{-4} \,,
\nen
that represent the probabilities to obtain by chance $\chi^2=0.035$ and $\chi^2=14.32$, with one and two degrees of freedom respectively, if the model is correct. This represents a very clear indication in favor of the hypothesis of two different ratios, $R_1$ and $R_2$. %
\begin{table}
\vspace{-2mm}
\centering
\caption{Input and output values of the BRs and their discrepancies, for the $\psii$ meson, obtained under the 2-$R$ hypothesis.}
\label{tab:BRcalc.psi2s} 
\begin{tabular}{lrrc} 
\hline\noalign{\smallskip}
\bb & BR$^{\rm exp}_{\bb}\times 10^4$ & BR$_{\bb}\times 10^4$ & Discr. $(\sigma)$ \\
\noalign{\smallskip}\hline\noalign{\smallskip}
$\Sigma^0 \overline \Sigma{}^0$ & $2.35 \pm 0.09$ & $2.35 \pm 0.16$ & $ 0 $ \\
$\Lambda \overline \Lambda$ & $3.81 \pm 0.13$ & $3.81 \pm 0.21$ & $ 0 $ \\
$\Lambda \overline \Sigma{}^0 + {\rm c.c.} $ & $0.120 \pm 0.024$ & $0.125 \pm 0.019$ & $ 0.088 $ \\
$p \overline p$ & $2.94 \pm 0.08$ & $2.94 \pm 0.23$ & $ 0 $ \\
$n \overline n$ & $3.06 \pm 0.15$ & $3.05 \pm 0.25$ & $ 0.066 $ \\
$\Sigma^+ \overline \Sigma{}^-$ & $2.32 \pm 0.12$ & $2.32 \pm 0.23$ & $ 0 $ \\
$\Xi^- \overline \Xi{}^+$ & $2.87 \pm 0.11$ & $2.86 \pm 0.23$ & $ 0.089 $ \\
$\Xi^0 \overline \Xi{}^0$ & $2.73 \pm 0.13$ & $2.74 \pm 0.21$ & $ 0.076 $ \\
$\gamma \to p \overline p$ & $0.0216 \pm 0.0019$ & $0.0216 \pm 0.0068$ & $ 0 $ \\
$\Sigma^- \overline \Sigma{}^+$ & / & $2.57 \pm 0.26$ & / \\
\noalign{\smallskip}\hline
\end{tabular}
\end{table}
\begin{figure}
\centering
	\includegraphics[width=.99\linewidth]{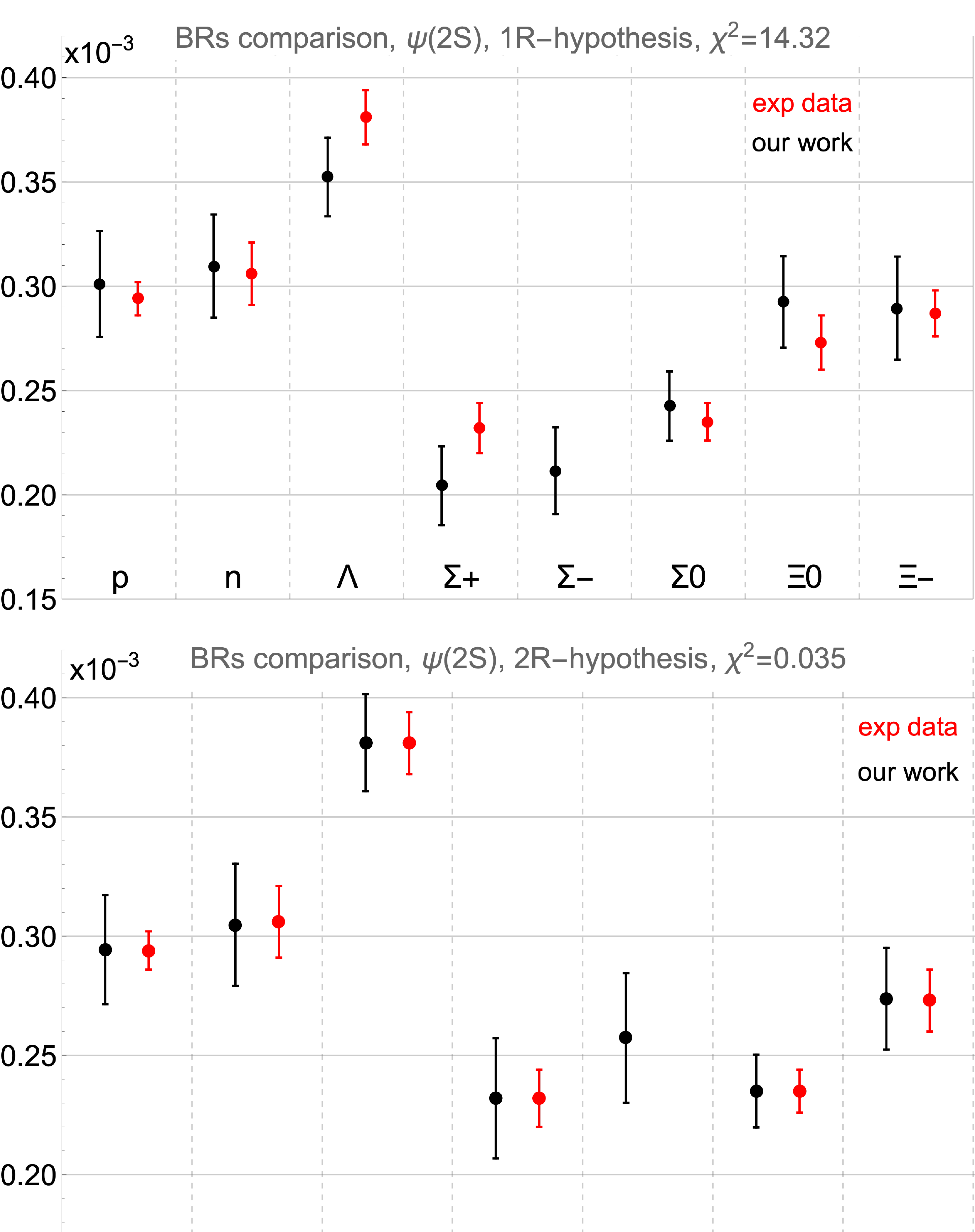}
\caption{\label{fig:br-comp} Comparison of BRs (experimental input vs model predictions) obtained under the 1-$R$ hypothesis (upper panel) and 2-$R$ hypothesis (lower panel). The red points are from Table~\ref{tab:BRJpsi.data}, while the black ones are the corresponding values obtained as outcomes of the minimization process. The errors are obtained by means of a Monte Carlo procedure.}
\end{figure}\\
Even though the $\chi^2$ is relatively high, the ratio given in Eq.~\eqref{eq-R.1R.S} could be considered as the representative mean value for the mixed-to-strong amplitude ratios for the decays of the $\psii$ meson into baryon pairs, under the 1-$R$ hypothesis. This value splits properly into $R_1$ and $R_2$ reported in the last two rows of Table~\ref{tab:results.par}. The value of Eq.~\eqref{eq-R.1R.S} should be compared with the corresponding one obtained for the $\jp$ meson, i.e., $R=-0.097 \pm 0.021$, see the last row of Table~\ref{tab:results.par-j}. The moduli of the two ratios differ from the asymptotic QCD prediction of Eq.~\eqref{eq:RpQCD}, i.e., $|R_{\rm pQCD}| \sim 0.030$. Naively, assuming a scaling proportional to the squared mass ratio, we obtain that starting from the energy squared 
\be
E^2_{\rm pQCD} \sim 22 \ \rm GeV^2 \,,
\nen
the QCD regime can be considered perturbative, as shown in Fig.~\ref{fig:epqcd}.
\begin{figure}
\centering
	\includegraphics[width=.99\linewidth]{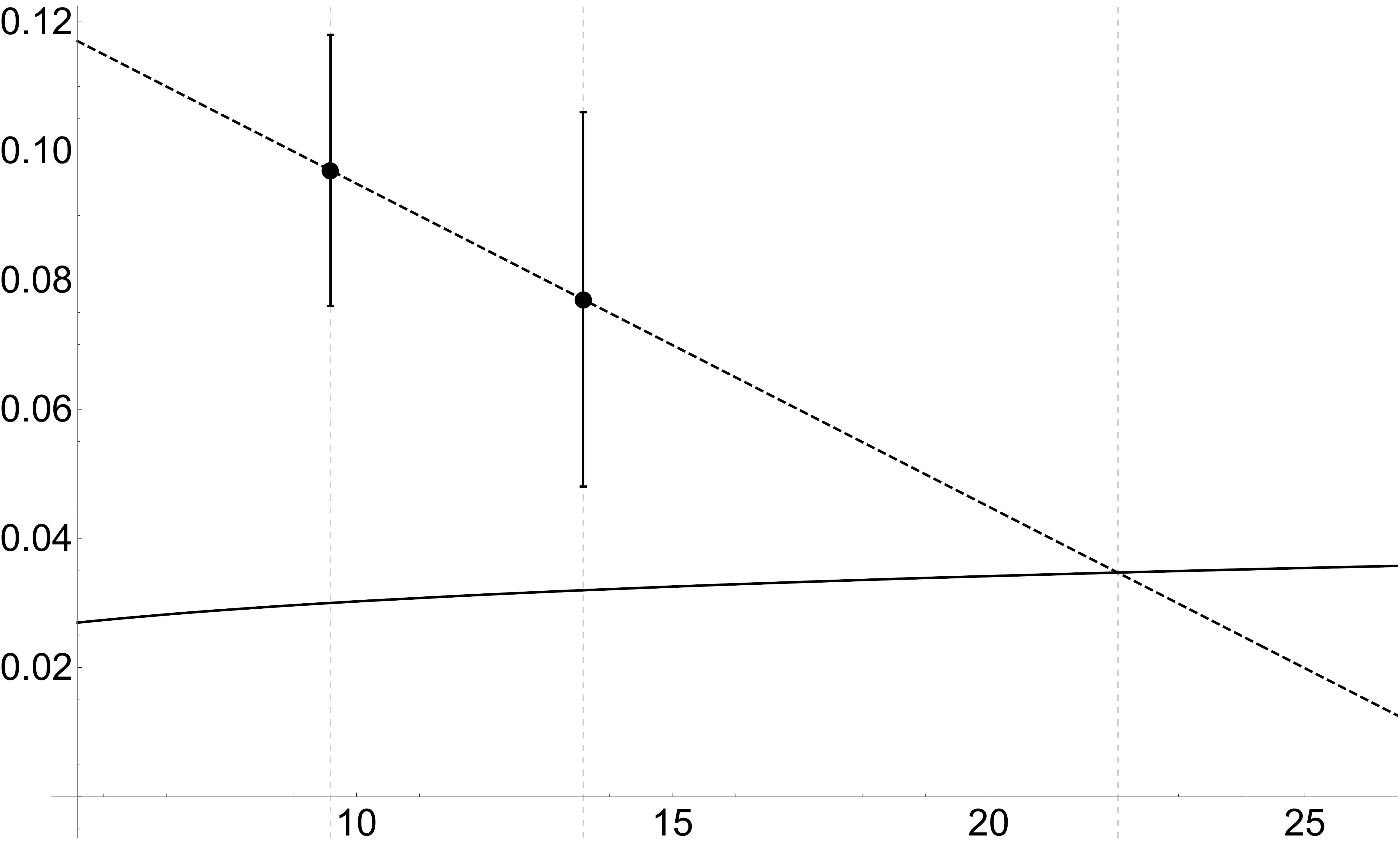}
	\put(-10,-8){$q^2$}
	\put(-236,147){$|R|$}
	\put(-187,-12){$M_{\jp}^2$}
	\put(-148,-12){$M_{\psii}^2$}
	\put(-58,-12){$E_{\rm pQCD}^2$}
	\put(-147,28){\small $|R_{\rm pQCD}|={4 \alpha \over 5 \alpha_S(q^2)}$}
	\caption{\label{fig:epqcd} An estimation of the energy region for the transition to the perturbative QCD regime. The black line represents the pQCD prediction for the modulus of the ratio defined in Eq.~\eqref{eq:RpQCD}.}
\end{figure}
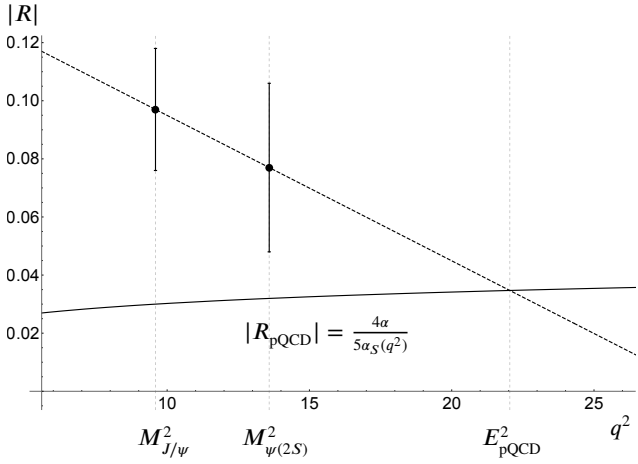\\
In the second and third and columns of Table~\ref{tab:BRcalc.psi2s} we report the input, output values of the BRs, while their discrepancy\footnote{As usually we define the discrepancy between two values with errors $x \pm \sigma_x$ and $y \pm \sigma_y$ as
\be
\mbox{Discr. $(\sigma)$} = {|x-y| \over \sqrt{\sigma_x^2+\sigma_y^2}}
\nen} is given in the fourth column.
\\
 A direct comparison between the BR discrepancies (input data vs model predictions) for the 1-$R$ and 2-$R$ hypotheses can be done by looking at Fig.~\ref{fig:br-comp}. It is clear that the 2-$R$ hypothesis is significantly better then the other one.
 \\
In Table~\ref{tab:BRsepcalc.psi2s} we report the single three contributions, purely strong, purely EM and mixed strong-EM, to the total BR. For a comparison, in Table~\ref{tab:BRsepcalc.j} we show the corresponding values obtained for the $\jp$ meson. %
\begin{table}
\vspace{-2mm}
\centering
\caption{Strong (second column), EM (third column) and mixed (fourth column) BRs for the $\psii$ meson under the 2-$R$ hypothesis.}
\label{tab:BRsepcalc.psi2s} 
\begin{tabular}{lrrr} 
\hline\noalign{\smallskip}
\bb & $\br^{ggg}_{\bb} \times 10^{4}$ & $\br^{\gamma}_{\bb} \times 10^{5}$ & $\br^{gg\gamma}_{\bb} \times 10^{5}$ \\
\noalign{\smallskip}\hline\noalign{\smallskip}%
$\Sigma^0 \overline \Sigma{}^0$            & $2.01 \pm 0.12$ & $0.41 \pm 0.79$       & $0$             \\
$\Lambda \overline \Lambda$              & $4.22 \pm 0.18$ & $0.43 \pm 0.81$       & $0$             \\
$\Lambda \overline \Sigma{}^0+$ c.c.  & $0$               & $1.25 \pm 0.24$         & $0$             \\
$p \overline p$                          & $3.74 \pm 0.14$ & $0.207 \pm 0.098$         & $0.90 \pm 0.33$ \\
$n \overline n$                          & $3.73 \pm 0.14$ & $1.85 \pm 0.35$         & $0$             \\
$\Sigma^+ \overline \Sigma{}^-$            & $2.02 \pm 0.12$ & $0.186 \pm 0.088$         & $0.043 \pm 0.059$ \\
$\Sigma^- \overline \Sigma{}^+$            & $2.01 \pm 0.12$ & $0.73 \pm 0.17$         & $0.044 \pm 0.060$ \\
$\Xi^- \overline \Xi{}^+$                  & $3.31 \pm 0.12$ & $0.67 \pm 0.16$         & $0.80 \pm 0.29$ \\
$\Xi^0 \overline \Xi{}^0$                  & $3.33 \pm 0.12$ & $1.50 \pm 0.29$         & $0$             \\
\noalign{\smallskip}\hline
\end{tabular}
\end{table} %
\begin{table}
\vspace{-2mm}
\centering
\caption{Strong (second column), EM (third column) and mixed (fourth column) BRs for the $\jp$ meson~\cite{Ferroli:2019nex} under the 1-$R$ hypothesis.}
\label{tab:BRsepcalc.j} 
\begin{tabular}{lrrr} 
\hline\noalign{\smallskip}
\bb & $\br^{ggg}_{\bb} \times 10^{3}$ & $\br^{\gamma}_{\bb} \times 10^{5}$ & $\br^{gg\gamma}_{\bb} \times 10^{5}$ \\
\noalign{\smallskip}\hline\noalign{\smallskip}%
$\Sigma^0 \overline \Sigma{}^0$            & $1.100 \pm 0.030$ & $0.902 \pm 0.076$       & $0$             \\
$\Lambda \overline \Lambda$              & $2.020 \pm 0.042$ & $0.981 \pm 0.083$       & $0$             \\
$\Lambda \overline \Sigma{}^0+$ c.c.  & $0$               & $2.83 \pm 0.24$         & $0$             \\
$p \overline p$                          & $2.220 \pm 0.085$ & $8.52 \pm 0.89$         & $2.19 \pm 0.93$ \\
$n \overline n$                          & $2.220 \pm 0.085$ & $4.50 \pm 0.38$         & $0$             \\
$\Sigma^+ \overline \Sigma{}^-$            & $1.100 \pm 0.030$ & $6.86 \pm 0.72$         & $1.08 \pm 0.46$ \\
$\Sigma^- \overline \Sigma{}^+$            & $1.090 \pm 0.030$ & $0.52 \pm 0.20$         & $1.07 \pm 0.46$ \\
$\Xi^- \overline \Xi{}^+$                  & $1.240 \pm 0.052$ & $0.43 \pm 0.16$         & $1.22 \pm 0.52$ \\
$\Xi^0 \overline \Xi{}^0$                  & $1.260 \pm 0.053$ & $2.99 \pm 0.25$         & $0$             \\
\noalign{\smallskip}\hline
\end{tabular}
\end{table}\\
A study of more general $3$-$R$ and $2$-$R$, with different subsets of final states, hypotheses is given in appendix~\ref{appA}. The results confirm that, in the case of the \psii\ meson, the optimal choice is that of the only two ratios, $R_1$ and $R_2$ given Eq.~\eqref{eq.choice.2R}. 
%
%
%
%
%
\section{$J/\psi$ results with 2-$R$ hypothesis}
\label{sec:resultsj}
We anticipate that the 1-$R$ hypothesis, i.e., $R=R_1=R_2$, is a good level-zero assumption for the $\jp$ decays. The obtained $\chi^2$ shown in Table~\ref{tab:results.par-j}, with 2 degrees of freedom, since there were 9 constraints and 7 free parameters, is a satisfactory result. 
We have applied the procedure based on the 2-$R$ hypothesis also to the $\jp$ meson. By minimizing the $\chi^2$ of Eq.~\eqref{eq.chi.2}, the best values for the parameters $G_0,D_e,D_m,F_e,F_m, \varphi$ are essentially unchanged compared to those obtained under the 1-$R$ hypothesis~\cite{Ferroli:2019nex}, reported in Table~\ref{tab:results.par-j}.
\\ 
The unexpected result concerns the values obtained for $R_1$ and $R_2$, being
\be
R_1 = -0.098 \pm 0.025 \,, \ \ \ \ \ R_2 = 0.07 \pm 0.11 \,,
\nen
with a minimum normalized $\chi^2$
\be
\frac{\chi^2\big(\xi^{\rm best}\big)}{N_{\rm dof}} = 0.00074\,,
\nen
to be compared, see Table~\ref{tab:results.par-j}, with the single ratio under the 1-$R$ hypothesis~\cite{Ferroli:2019nex}, i.e., 
\be
R = -0.097 \pm 0.021 \,,
\nen
with a minimum normalized $\chi^2$
\be
\frac{\chi^2\big({\xi'}^{\rm best}\big)}{N'_{\rm dof}} = 1.33 \,.
\nen
We observed a difference in the ratios for the charged sigma baryons, as for the $\psii$ meson, but in a less evident way, considering the $\chi^2$ values and the errors obtained in the various cases.\\
 In particular, also for the $\jp$ meson, we estimate the significance of the different mixed-to-strong amplitude ratios by considering the following $p$-values
\be
p(2.65;2)= 0.266 \,, \ \ \ \ \ p(0.00074;1)= 0.979 \,,
\nen
corresponding to the 1-$R$ and 2-$R$ hypotheses. These values confirm the hypothesis, more evident in the $\psii$ case, of a different behavior for the decays into charged sigma baryon. 
\\
The only difference with respect to the results obtained under the 1-$R$ hypothesis, see Table~\ref{tab:BRsepcalc.j}, is given by the BRs  
\be
\begin{aligned}
\br^{gg\gamma}_{\Sigma^+ \overline \Sigma{}^-} &= (1.9 \pm 2.6) \times 10^{-5} \,,\\
\br^{gg\gamma}_{\Sigma^- \overline \Sigma{}^+} &= (2.0 \pm 2.6) \times 10^{-5} \,.
\label{eq.br.mix.s-j}
\end{aligned}
\en
\section{Conclusions}
The study of the $\psii$ meson suggests a different behavior in the decays into spin 1/2 baryons with respect to the $\jp$ meson~\cite{Ferroli:2019nex}. In particular, for the $\jp$ meson the $\chi^2$ minimization is satisfying with a parametrization of the decay amplitudes in terms of seven the parameters $G_0$, $D_e$, $D_m$, $F_e$, $F_m$, $R$, $\varphi$, see Table~\ref{tab:A.par.pre} with $R=R_1=R_2$, and nine constraints, see Table~\ref{tab:results.par-j}.
\\
The same procedure applied to the $\psii$ meson gives worst results with a relatively larger $\chi^2$, see Eq.~\eqref{eq:chi2-2} and Fig.~\ref{fig:br-comp}. As consequence we introduced an additional degree of freedom to separate the mixed-to-strong amplitude ratio in two parameters $R_1$ for $p \overline p$, $\Xi^- \overline \Xi{}^+$, and $R_2$ for $\Sigma^+ \overline \Sigma{}^-$, $\Sigma^- \overline \Sigma{}^+$.
\\
This choice is also supported by the results obtained under more general hypotheses investigated in appendix~\ref{appA}.
\\
The new parametrization, detailed in Table~\ref{tab:A.par.pre}, leads to a satisfactory minimization of the $\chi^2$, whose results are reported in Table~\ref{tab:results.par}.
\\
 The application of this new approach to the \jp\ meson gave results almost compatible with those of Table~\ref{tab:results.par-j}, with the obvious  exception of the new parameters $R_1$ and $R_2$. 
 \\
 The obtained $R_1$ and the $R$ value of Table~\ref{tab:results.par-j} are quite similar, on the contrast $R_2$ is also compatible with zero. These results suggest a different behavior of the charged sigma baryons, as for the $\psii$ meson, where the difference is significantly more evident.
\\
By comparing the obtained parameters for the $\psii$, see Table~\ref{tab:results.par}, with those of the $\jp$, see Table~\ref{tab:results.par-j}, we observe that the SU(3) symmetry breaking, represented by $D_m$ and $F_m$, plays a different role in the two cases. In particular, these parameters have the same sign in the case of the $\psii$, while they have different sign in the case of the $\jp$. Similar differences, due to the SU(3) breaking terms, have also been observed by studying the angular distributions of the decays into lambda and sigma baryons~\cite{Ablikim:2017tys,Ferroli:2018yad}.
\\
The BRs obtained by the $\chi^2$ minimization procedure are fully in agreement with the corresponding input values, see Table~\ref{tab:BRcalc.psi2s} and Fig.~\ref{fig:br-comp} (lower panel). Moreover, the BR value $\br_{\Sigma^- \overline \Sigma{}^+} = (2.57 \pm 0.26) \times 10^{-4}$ represents a prediction of the model, since there are no available data.
\\
In the \psii\ case, we obtained a strong-EM relative phase of $\varphi = (58\pm 8)^\circ$ and the confirmation that at this energy range the QCD regime is still not completely perturbative. Moreover, by using both $\jp$ and $\psii$ results we can identify with $E^2_{\rm p QCD} \sim 22$ GeV$^2$ the energy region where the QCD becomes perturbative, see Figure~\ref{fig:epqcd}.
\\
As done for the $\jp$ meson, we separated, for the first time, the strong, the EM and the mixed strong-EM contributions to the total BR of the decays $\psii\to\bb$. The obtained values, see Table~\ref{tab:BRsepcalc.psi2s}, can be compared to those of the $\jp$ meson, see Table~\ref{tab:BRsepcalc.j}. 
\\
It is evident that the purely strong contribution to the total BR is predominant in the case of the $\jp$ with respect to the $\psii$. In particular, the ratios of the EM and the strong contributions to the total BR are of order $10^{-1}$ for the $\psii$ and $10^{-2}$ for the $\jp$. Such ratios are compatible with the decreasing trend suggested by the pQCD, when $q^2\gg\Lambda^2_{\rm QCD}$.
\\
 In the case of the $\psii$ meson, the mixed strong-EM contributions, see Table~\ref{tab:BRsepcalc.psi2s}, are different for the two pairs of final states: $p \overline p$, $\Xi^- \overline \Xi{}^+$ and $\Sigma^+ \overline \Sigma{}^-$, $\Sigma^- \overline \Sigma{}^+$. In particular, for the first pair they are of the same order as the EM contributions. On the other hand, for the charged sigma baryons, they are at least one order of magnitude lower and compatible with zero. A similar trend can be inferred, in a less evident way, also by looking at the values of Eq.~\eqref{eq.br.mix.s-j}, which are compatible with the corresponding values given in the Table~\ref{tab:BRsepcalc.j}, under the 1-$R$ hypothesis.
\section*{Acknowledgement}
This work was supported in part by the STRONG-2020 project of the European Union's Horizon 2020 research and innovation programme under grant agreement No.~824093.
%
%
%
\appendix
\section{Analysis of the $n$-$R$ hypotheses}
\label{appA}
The parameter $R_\b={\mathcal{A}_{\bb}^{gg\gamma}}/{\mathcal{A}_{\bb}^{ggg}}$, i.e., the ratio between the mixed strong-EM and the strong amplitudes of the decay $\psii\to\bb$, is defined only for the charged baryons, being $\mathcal{A}_{\bb}^{gg\gamma}=0$ for neutral baryons. 
\\
As a consequence, in the most general case, that is by considering the lower degree of degeneracy, we would have a maximum of three ratios, namely: $R_{p}$, $R_{\Xi}$ and $R_{\Sigma}$, one for each charged-baryon species.
\\
In case of degeneracy, instead, two ratios or even only one common ratio can be considered. To summarize, in terms of the degree of degeneracy, we have:
\begin{itemize}
\item no degeneracy, this is the so-called 3-$R$ hypothesis corresponding the single case 
\be
R_1 = R_{p} \neq R_2=R_{\Sigma} \neq R_3=R_{\Xi}\,;
\nen
\item minimum degeneracy, 2-$R$ hypothesis, under which there are three cases, corresponding to the three possible identification of two out of three ratios,
\be
R_1 \ug R_{p} =R_{\Sigma}\neq R_2=R_{\Xi}\,,\no\\
R_1 \ug R_{\Sigma} =R_{\Xi} \neq R_2= R_{p}\,,\no\\
R_1 \ug R_{\Xi}=R_{p} \neq R_2=R_{\Sigma}\,,
\nen
the last case, where the proton and the cascade have a common ratio, is the most favored by the data;
\item maximum degeneracy, 1-$R$ hypothesis, that, as in the case of no degeneracy, entails only one possibility, i.e.,
\be
R_1 \ug R_{p} =R_{\Sigma}=R_{\Xi}\,.
\nen
\end{itemize}
Tables~\ref{tab:allR2S} and~\ref{tab:allRJ} report the results for the ratios as well as the $\chi^2$'s obtained in all the cases. 
\\
It is interesting to notice how, for the \psii\ meson, the association $R_1 = R_{\Xi}=R_{p}$, under the 2-$R$ hypothesis (third case in Table~\ref{tab:allR2S}), appears as the most favored, not only in the light of its lower $\chi^2$, but also because such an association is phenomenologically supported by the results of the most general non-degeneracy case.
\begin{table}
\vspace{-2mm}
\centering
\caption{Comparison of the $\chi^2$ values obtained under the hypotheses of 3-$R$ (one case), 2-$R$ (three cases) and 1-$R$ (one case) for the $\psii$ meson. The symbol $\dagger$ indicates the ratios that have a common value in the three cases considered under the 2-$R$ hypothesis.}
\label{tab:allR2S} 
\begin{tabular}{lrrrc} 
\hline\noalign{\smallskip}
$R_{p}$ & $R_{\Sigma}$ & $R_{\Xi}$ & $\chi^2$ & hypoth. \\
\noalign{\smallskip}\hline\noalign{\smallskip}
\multicolumn{3}{c}{$-0.077(29)$} & $14.32$ & 1-$R$ \\
\noalign{\smallskip}\hline\noalign{\smallskip}
$-0.07(3)^\dagger$ & $-0.07(3)^\dagger$ & $-0.09(4)$ & $14.07$ & \multirow{3}{*}{2-$R$}\\
$-0.15(4)$ & $-0.06(3)^\dagger$ & $-0.06(3)^\dagger$ & $8.04$ & \\
$-0.15(3)^\dagger$ & $0.021(50)$ & $-0.15(3)^\dagger$ & $0.035$ &  \\
\noalign{\smallskip}\hline\noalign{\smallskip}
$-0.16(3)$ & $0.018(48)$ & $-0.15(4)$ & $6 \times 10^{-7}$ & 3-$R$ \\
\noalign{\smallskip}\hline
\end{tabular}
\end{table}%
\begin{table}
\vspace{-2mm}
\centering
\caption{Comparison of the $\chi^2$ values obtained under the hypotheses of 3-$R$ (one case), 2-$R$ (three cases) and 1-$R$ (one case) for the $\jp$ meson. The symbol $\dagger$ indicates the ratios that have a common value in the three cases considered under the 2-$R$ hypothesis.}
\label{tab:allRJ} 
\begin{tabular}{lrrrc} 
\hline\noalign{\smallskip}
$R_{p}$ & $R_{\Sigma}$ & $R_{\Xi}$ & $\chi^2$ & hypoth. \\
\noalign{\smallskip}\hline\noalign{\smallskip}
\multicolumn{3}{c}{$-0.097(21)$} & $2.66$ & 1-$R$ \\
\noalign{\smallskip}\hline\noalign{\smallskip}
$-0.10(3)^\dagger$ & $-0.10(3)^\dagger$ & $-0.10(4)$ & $2.64$ & \multirow{3}{*}{2-$R$}\\
$-0.11(3)$ & $-0.08(4)^\dagger$ & $-0.08(4)^\dagger$ & $2.35$ & \\
$-0.10(2)^\dagger$ & $0.07(11)$ & $-0.10(2)^\dagger$ & $7 \times 10^{-4}$ &  \\\noalign{\smallskip}\hline\noalign{\smallskip}
$-0.10(3)$ & $0.07(10)$ & $-0.10(5)$ & $6 \times 10^{-4}$ & 3-$R$ \\
\noalign{\smallskip}\hline
\end{tabular}
\end{table}%
%

%
%
%

%
%
%
%

\begin{thebibliography}{}
%
%
\bibitem{Tanabashi:2018oca}
M.~Tanabashi \textit{et al.} [Particle Data Group],
Phys. Rev. D \textbf{98} (2018) no.3, 030001
doi:10.1103/PhysRevD.98.030001

\bibitem{Ferroli:2019nex}
R.~Baldini Ferroli, A.~Mangoni, S.~Pacetti and K.~Zhu,
Phys. Lett. B \textbf{799} (2019), 135041
doi:10.1016/j.physletb.2019.135041
[arXiv:1905.01069 [hep-ph]].
\bibitem{Ferroli:2016jri}
R.~Baldini Ferroli, A.~Mangoni and S.~Pacetti,
Phys. Rev. C \textbf{98} (2018) no.4, 045210
doi:10.1103/PhysRevC.98.045210
[arXiv:1611.04437 [hep-ph]].

\bibitem{Ablikim:2017tys}
M.~Ablikim \textit{et al.} [BESIII],
Phys. Rev. D \textbf{95} (2017) no.5, 052003
doi:10.1103/PhysRevD.95.052003
[arXiv:1701.07191 [hep-ex]].

%
\bibitem{Ferroli:2018yad}
M.~Alekseev, A.~Amoroso, R.~B.~Ferroli, I.~Balossino, M.~Bertani, D.~Bettoni, F.~Bianchi, J.~Chai, G.~Cibinetto, F.~Cossio, F.~Mori, M.~Destefanis, R.~Farinelli, L.~Fava, G.~Felici, I.~Garzia, M.~Greco, L.~Lavezzi, C.~Leng, M.~Maggiora, A.~Mangoni, S.~Marcello, G.~Mezzadri, S.~Pacetti, P.~Patteri, A.~Rivetti, M.~Da Rocha Rolo, M.~Savrié, S.~Sosio, S.~Spataro and L.~Yan,
Chin. Phys. C \textbf{43} (2019) no.2, 023103
doi:10.1088/1674-1137/43/2/023103
[arXiv:1809.04273 [hep-ph]].
%

\bibitem{Zhu:2015bha}
K.~Zhu, X.~H.~Mo and C.~Z.~Yuan,
Int. J. Mod. Phys. A \textbf{30} (2015) no.25, 1550148
doi:10.1142/S0217751X15501481
[arXiv:1505.03930 [hep-ph]].

\bibitem{Korner:1986vi}
J.~G.~Korner,
Z. Phys. C \textbf{33} (1987), 529
doi:10.1007/BF01548265

\bibitem{Chernyak:1984bm}
V.~Chernyak and I.~Zhitnitsky,
Nucl. Phys. B \textbf{246} (1984), 52-74
doi:10.1016/0550-3213(84)90114-7

\bibitem{Claudson:1981fj}
M.~Claudson, S.~L.~Glashow and M.~B.~Wise,
Phys. Rev. D \textbf{25} (1982), 1345
doi:10.1103/PhysRevD.25.1345

\bibitem{Ablikim:2019njl}
M.~Ablikim \textit{et al.} [BESIII],
Phys. Rev. D \textbf{99} (2019) no.9, 092002
doi:10.1103/PhysRevD.99.092002
[arXiv:1902.00665 [hep-ex]].

\bibitem{Ablikim:2016sjb}
M.~Ablikim \textit{et al.} [BESIII],
Phys. Lett. B \textbf{770} (2017), 217-225
doi:10.1016/j.physletb.2017.04.048
[arXiv:1612.08664 [hep-ex]].

%
\end{thebibliography}

\end{document}